\documentstyle[aps,pre,epsfig,multicol]{revtex}

\begin{document}
\draft

\title{Anderson localization as a parametric instability of the
linear kicked oscillator}

\author{L.~Tessieri and F.~M.~Izrailev}

\address{
Instituto de F\'{\i}sica, Universidad Aut\'{o}noma de Puebla, \\
Apdo. Postal J-48, Puebla, Pue. 72570, M\'{e}xico}

\date{14th December 1999}

\maketitle

\begin{abstract}
We rigorously analyse the correspondence between the
one-dimensional standard Anderson model and a related classical
system, the `kicked oscillator' with noisy frequency. We show that
the Anderson localization corresponds to a parametric instability
of the oscillator, with the localization length determined by an
increment of the exponential growth of the energy. Analytical
expression for a weak disorder is obtained, which is valid both
inside the energy band and at the band edge.
\end{abstract}


\begin{multicols}{2}
\section{Introduction}

Recently it was shown that quantum one-dimensional tight-binding
models with any diagonal site-potential can be formally represented
in terms of a two-dimensional Hamiltonian map~\cite{Izr95}.
On the other hand, this classical map is associated with a linear
oscillator subjected to a linear force given in the form of
time-dependent delta-kicks. In this picture, both the frequency of
the unperturbed oscillator and the period of the kicks are determined
by the energy of an eigenstate, and the amplitudes of the kicks are
defined by the site-potential in the original quantum model.
It was shown that by exploring the dynamics of this classical system,
one can obtain global characteristics of the eigenstates, such as the
localization length defined by the Lyapunov exponent of classical
trajectories.

In particular, analytical estimates have been obtained in~\cite{Izr95}
for a specific diagonal site-potential potential with short-range
correlations (the so-called dimer model, see~\cite{Dun90}). Other
applications to the case of general correlated diagonal~\cite{Izr99}
and off-diagonal~\cite{Tes99} disorder have revealed very important
peculiarities. One of the most interesting results has been obtained
in~\cite{Izr99} where it was shown how to construct random potentials
with specific two-point correlators which result in the emergence of
the mobility edge in one-dimensional geometry. Based on these
predictions, very recently experimental realization of this effect
has been done in one-mode microwave guides~\cite{Kuh99}.

In this paper, we perform an analytical study of the standard
Anderson model with diagonal uncorrelated disorder, paying main
attention to the problem of the mathematical correspondence
between the quantum model and its classical representation in the
form of a linear kicked oscillator. More specifically, we are
interested in the connection between the Anderson localization and
the paramtric instability of the corresponding classical system.
Although the results obtained for the localization length in the
weak disorder limit are well known from other studies, the method
we use here is a new one and it may explain the mechanism of the
Anderson transition in new terms. Moreover, this approach may be
very useful for 2D and 3D cases, for which analytical results for
global properties of eigenstates are very restricted.

\section{Definition of the model}

In this paper we study the relation existing between the standard 1D
Anderson model and a related physical system, a linear oscillator with
noisy frequency. The quantum model is defined by the stationary
Schr\"{o}dinger equation~\cite{And58}
\begin{equation}
\psi_{n+1} + \psi_{n-1} + \epsilon_{n} \psi_{n} = E \psi_{n},
\label{andmod}
\end{equation}
where $\psi_{n}$ represents the electron wave-function at
the $n$-th lattice site, and the site-energies $\epsilon_{n}$ are
independent random variables with a common distribution $p(\epsilon)$.
In the standard Anderson model the probability $p(\epsilon)$ has the
form of a box distribution,
\begin{equation}
p (\epsilon) = \frac{1}{W} \theta \left( \frac{W}{2} - |\epsilon|
\right),
\label{boxdis}
\end{equation}
whose width $W$ sets the strength of the disorder. In the following,
however, we will not restrict our considerations to the specific
form~(\ref{boxdis}) of the probability distribution, but simply assume
that the variables $\epsilon_{n}$ have zero mean
($\langle \epsilon_{n} \rangle = 0$) and a finite variance $\langle
\epsilon_{n}^{2} \rangle$.

The kicked oscillator is a harmonic oscillator that undergoes periodic
and instantaneous variations of the momentum (the `kicks'). Such a system
is defined by the Hamiltonian
\begin{equation}
H = \omega \left( \frac{p^2}{2} + \frac{x^2}{2} \right) +
\frac{x^2}{2} \xi (t),
\label{kickosc}
\end{equation}
where
\begin{equation}
\xi \left( t \right) = \sum_{n = -\infty}^{+\infty} A_{n} \; \delta
\left( t - n T \right) .
\label{xi}
\end{equation}
The random coefficients $A_{n}$ that appear in the definition of the
noise~(\ref{xi}) represent the intensity of the `kicks', i.e., they are
proportional to the sudden momentum changes experienced by the oscillator
at times $t = n T$.
In other words, the system~(\ref{kickosc}) represents a harmonic oscillator
with a mean frequency $\omega$ perturbed by the noise term  $\xi(t)$.
Using the definition~(\ref{xi}), one can easily reconduct the statistical
properties of the noise $\xi(t)$ to the corresponding properties of the
variables $A_{n}$; in particular, the mean and the variance of $\xi(t)$ can
be expressed as
\begin{eqnarray*}
\langle \xi (t) \rangle = \lim_{\tau \to \infty} \frac{1}{\tau}
\int_{-\tau/2}^{\tau/2} \xi(t) \; dt \\
= \lim_{N \to \infty} \frac{1}{NT}
\sum_{n=-N/2}^{N/2} A_{n} = \frac{\langle A_{n} \rangle}{T}
\end{eqnarray*}
and
\begin{eqnarray*}
\langle \xi(t) \xi(t+s) \rangle = \lim_{\tau \to \infty}
\frac{1}{\tau} \int_{-\tau/2}^{\tau/2} \xi(t) \xi(t+s) \; dt \\
= \frac{1}{T} \delta \left( s \right) \;
\lim_{N \to \infty} \frac{1}{N} \sum_{n=-N/2}^{N/2} A_{n}^{2} =
\frac{\langle A_{n}^{2} \rangle}{T} \; \delta \left( s \right) .
\end{eqnarray*}

The equivalence of the models~(\ref{andmod}) and~(\ref{kickosc}) has been
discussed in~\cite{Izr95}. There it was shown how the two-dimensional map
\begin{equation}
\left\{
\begin{array}{ccc}
x_{n+1} & = & x_{n} \cos \left( \omega T \right) + \left( p_{n} - A_{n}
x_{n} \right) \sin \left( \omega T \right) \\
p_{n+1} & = & - x_{n} \sin \left( \omega T \right) + \left( p_{n} - A_{n}
x_{n} \right) \cos \left( \omega T \right)
\end{array}
\right.
\label{map}
\end{equation}
can be derived by integrating over a period $T$ the Hamiltonian equations
of motion of the kicked oscillator~(\ref{kickosc}). Note that in the
map~(\ref{map}) $x_{n}$ and $p_{n}$ stand for the coordinate and momentum of
the oscillator immediately before the $n$-th kick.
Eliminating the momentum from equations~(\ref{map}), one eventually
obtains the relation
\begin{displaymath}
x_{n+1} + x_{n-1} + A_{n} \cos \left( \omega T \right) x_{n} =
2 x_{n} \cos \left( \omega T \right) ,
\end{displaymath}
which coincides with the Anderson equation~(\ref{andmod}) if one
identifies the site amplitude $\psi_{n}$ with the coordinate $x_{n}$ of the
oscillator and if the parameters of the models~(\ref{andmod})
and~(\ref{kickosc}) are related to each other by the equalities
\begin{eqnarray}
\epsilon_{n} = A_{n} \sin \left( \omega T \right) & \mbox{;} \;\;\;\; &
E = 2 \cos \left( \omega T \right) .
\label{param}
\end{eqnarray}

In Ref.~\cite{Izr95} the classical map~(\ref{map}) was used as a tool
to investigate the properties of the model~(\ref{andmod}); here we
focus instead on the direct analysis of the Hamiltonian
model~(\ref{kickosc}).

\section{The oscillator with noisy frequency}

The dynamics of the kicked oscillator~(\ref{kickosc}) is determined by the
Hamiltonian equations of motion:
\begin{equation}
\left\{
\begin{array}{ccl}
\dot{p} & = & - \left( \omega + \xi \left( t \right) \right) x \\
\dot{x} & = & \omega p
\end{array} .
\right.
\label{dyneq}
\end{equation}
In order to study the behaviour of the kicked oscillator, it is convenient
to substitute the couple of differential equations~(\ref{dyneq}) with
the system of stochastic It\^{o} equations,
\begin{equation}
\left\{
\begin{array}{ccl}
dp & = & - \omega x \; dt - x \sqrt{\langle A_{n}^{2} \rangle/T}
\; dW(t) \\
dx & = & \omega p \; dt
\end{array}
\right.
\label{itoeq}
\end{equation}
where $W(t)$ is a Wiener process with $\langle dW(t) \rangle = 0$ and
$\langle dW(t)^{2} \rangle = dt$.

The systems~(\ref{dyneq}) and~(\ref{itoeq}) can be considered equivalent
inasmuch as the shot noise $\xi(t)$ is adequately represented by
a Wiener process $W(t)$. That is the case if the strength of the single
kicks is weak, i.e., if the condition
\begin{equation}
\langle A_{n}^{2} \rangle \ll 1
\label{weakdis}
\end{equation}
is fulfilled.
That can be understood by considering that the present situation is
analogous to the one that occurs in the Brownian motion of a heavy
particle suspended in a fluid of light molecules.
The instantaneous impacts of the fluid molecules on the massive particle
can be successfully described by a continuous Wiener process, provided that
each single collision does not produce a significant displacement of the
heavy particle.
When the mass of the suspended particle is not much bigger than the one of
the impinging molecules, the nature of the motion changes and the effect
of the molecular collisions can no longer be depicted by a Wiener process.

A similar analogy can be drawn between the present case and the
random walk problem. To be exact, let us consider a one-dimensional
random walk made by someone that takes steps of length $l$ at times $n T$
(with $n$ integral). At each step the walker is supposed to go to the
right or to the left with equal probability.
In this model the walker's position changes with each step much in the
way the momentum of the kicked oscillator does under the action of a kick:
in both cases the relevant physical quantity is varied in a sudden and
random way at regular time intervals.
This analogy makes interesting to notice that, by going to the limit
\begin{displaymath}
l \rightarrow 0, \;\;\; T \rightarrow 0 ,
\end{displaymath}
while holding fixed the ratio
\begin{displaymath}
D = \frac{l^{2}}{T}
\end{displaymath}
the discrete time random walk evolves in a Brownian motion with diffusion
constant $D$ (see, e.g., Ref.~\cite{Gar83}). In other words, a Wiener
process can be regarded as a limit case of random walk in the limit of
very small and fast-spaced steps.
In a similar way, the `jump process' $\xi(t)$ can be described by a
`diffusion process' $W(t)$ when the condition~(\ref{weakdis}) is satisfied,
with the ratio
\begin{equation}
k = \frac{\langle A_{n}^{2} \rangle}{\omega T}
\label{kappa}
\end{equation}
playing a role analogous to that of the diffusion constant $D$.

This conclusion, although substantially correct, must be made more
precise.
The analogy between the kicked oscillator~(\ref{kickosc}) and the free
Brownian particle or the random walker, although of much use, cannot
be complete because the kicked oscillator, unlike the two latter systems,
is endowed with an autonomous dynamics dictated by the elastic force and
independent from the noise.
As a consequence, in assessing the equivalence of the shot noise $\xi(t)$
with the continuous process $W(t)$, one must also take into account the
possibility that the interplay of the discrete noise~(\ref{xi}) with the
unperturbed motion of the oscillator might produce different effects than
those induced by the addition of the continuous process $W(t)$ to the
deterministic dynamics of the noiseless oscillator.
More specifically, one can guess that possible `resonance' effects
due to the commensurability of the frequency $1/T$ of the kicks with
the frequency $\omega/2\pi$ of the unperturbed oscillator might
occur. This is actually the case at the band centre, when the two
frequencies stand in the ratio $\omega T/2\pi = 1/4$, as we will discuss
in Sec.~\ref{wde}.
Apart from this exceptional case, however, the dynamical features of
the oscillators~(\ref{kickosc}) and~(\ref{itoeq}) do not differ, as will
appear in the following analysis.

In conclusion, the kicked oscillator~(\ref{dyneq}) and the stochastic
oscillator~(\ref{itoeq}) are equivalent when the individual
kicks of the first model are weak. On the other hand, since the kicked
oscillator~(\ref{dyneq}) is equivalent to the Anderson
model~(\ref{andmod}), one can conclude that the stochastic
oscillator~(\ref{itoeq}) provides a correct description of the Anderson
model for the weak disorder case.
This allows one to analyse the solutions of Eq.~(\ref{andmod}) in terms
of phase-space `trajectories' of the stochastic oscillator~(\ref{itoeq}).
In this picture, localized states for the Anderson model correspond
to unbounded trajectories of the oscillator in the phase space, while
extended states translate into bounded trajectories.

In the following we will restrict our considerations to the case defined
by the weak disorder condition~(\ref{weakdis}) and thus ensure the
equivalence between the Anderson model~(\ref{andmod}) and the stochastic
oscillator~(\ref{itoeq}).

\section{Lyapunov exponent}

Once we have established the correspondence of the Anderson model with
the stochastic oscillator~(\ref{itoeq}), we can proceed to redefine
essential features of the first model in the dynamical language of the
second. In particular, we are interested in deriving a formula for the
Lyapunov exponent, which gives the inverse localization length for the
eigenstates of the equation~(\ref{andmod}). Since these eigenstates
correspond to trajectories of the stochastic oscillator~(\ref{itoeq}),
the Lyapunov exponent is naturally redefined as the exponential
divergence rate of neighbouring trajectories, i.e. through the limit
\begin{equation}
\lambda = \lim_{T \to \infty} \; \lim_{\delta \to 0}
\frac{1}{T} \int_{0}^{T} dt \;\; \frac{1}{\delta} \log
\frac{x(t+\delta)}{x(t)} ,
\label{divrate}
\end{equation}
which corresponds to the standard expression
\begin{displaymath}
\lambda = \lim_{N \rightarrow \infty} \frac{1}{N} \sum_{n=1}^{N}
\log \frac{\psi_{n+1}}{\psi_{n}}
\end{displaymath}
for the Anderson model~(\ref{andmod}).
By taking the limit $\delta \to 0$ first, the expression~(\ref{divrate})
can be put in the simpler form
\begin{equation}
\lambda = \langle z \rangle = \lim_{T \to \infty} \frac{1}{T}
\int_{0}^{T} dt \;\; z(t) ,
\label{lyapdef}
\end{equation}
where the Ricatti variable $z = \dot{x}/x$ has been introduced and
the symbol $\langle z \rangle$ stands for the (time) average of $z$.

To compute the Lyapunov exponent, as defined by Eq.~(\ref{lyapdef}), it
is necessary to analyse the dynamics of the variable $z = \omega p/x$.
The time evolution of this quantity is determined by the It\^{o}
stochastic equation
\begin{equation}
dz = -\left( \omega^{2} + z^{2} \right) dt - \omega \sqrt{ \langle
A_{n}^{2} \rangle/T} \; dW(t) \; ,
\label{dz}
\end{equation}
which can be easily derived using~(\ref{itoeq}) and the standard rules
of the It\^{o} calculus.

Notice that, while the position and momentum of the
oscillator~(\ref{itoeq}) do not evolve independently from each other,
the dynamics of their ratio $z =\omega p/x$ is totally autonomous from
that of any other variable. As a consequence, one deals with the single
differential equation~(\ref{dz}) instead of having to cope with a set
of coupled equations like~(\ref{itoeq}).
Thus, the introduction of the variable $z$, which is suggested by the
definition~(\ref{lyapdef}) of the Lyapunov exponent, turns out to be
beneficial also for the study of the stochastic oscillator~(\ref{itoeq}).

As is known, the It\^{o} stochastic differential equation~(\ref{dz}) is
equivalent to the Fokker-Planck equation~\cite{Gar83,Kam92}
\begin{equation}
\frac{\partial p}{\partial t} (z,t) = \frac{\partial}{\partial z}
\left[ \left( \omega^{2} + z^{2} \right) p(z,t) \right] +
\frac{\omega^{2} \langle A_{n}^{2} \rangle}{2T} \frac{\partial^{2} p}
{\partial z^{2}}(z,t) \; ,
\label{fokpl}
\end{equation}
which gives the time evolution of the probability density $p(z,t)$ of the
stochastic variable $z$.
In other words, the evolution of $z(t)$ dictated by Eq.~(\ref{dz}) is a
diffusion process with a deterministic drift coefficient $\left( \omega^{2}
+ z^{2} \right)$ and a noise-induced diffusion coefficient $\omega^{2}
\langle A_{n}^{2} \rangle/T$.

The stationary solution of Eq.~(\ref{fokpl}) is
\begin{displaymath}
p(z) = \left[ C_{1} + C_{2} \int_{-\infty}^{z} dx \; \exp  \left\{
\Phi(x/\omega) \right\} \right] \, \exp \left\{ -\Phi(z/\omega) \right\} \; ,
\end{displaymath}
where $C_{1}$ and $C_{2}$ are integration constants and the function
$\Phi(x)$ is given by the relation
\begin{equation}
\Phi (x) = \frac{2}{k} \left( x + \frac{x^{3}}{3}   \right) \; ,
\label{phi}
\end{equation}
which contains the parameter $k$ defined by Eq.~(\ref{kappa}).
Since $p(z)$ is a probability distribution, it must be integrable and
therefore the constant $C_{1}$ must vanish. The residual constant $C_{2}$
is determined by the normalization condition $\int_{-\infty}^{\infty}
p(z) \; dz = 1$.
The resulting distribution is:
\begin{equation}
p(z) = \frac{1}{N \omega^{2}}
\int_{-\infty}^{z} dx  \; \exp \left\{ \Phi \left( x/\omega \right)
- \Phi \left( z/\omega \right) \right\}
\label{misinv}
\end{equation}
with
\begin{equation}
N = \sqrt{\frac{\pi k}{2}}
\int_{0}^{\infty} dx \; \frac{1}{\sqrt{x}} \exp \left[ -\frac{2}{k}
\left( x + \frac{x^{3}}{12} \right) \right] \;.
\label{norm}
\end{equation}

Once the steady-state probability distribution~(\ref{misinv}) is known,
one can use it to compute the average of $z$ that defines the Lyapunov
exponent~(\ref{lyapdef})
\begin{equation}
\lambda = \langle z \rangle = \int_{-\infty}^{\infty} dz \; z \, p(z).
\label{averag}
\end{equation}
By this way one obtains
\begin{equation}
\lambda = \frac{\omega}{2N} \int_{0}^{\infty} dx \; \sqrt{x} \exp
\left[ -\frac{2}{k} \left( x + \frac{x^{3}}{12} \right) \right] \;.
\label{lyap}
\end{equation}
Formula~(\ref{lyap}) is the central result of this paper. It gives an
expression for the inverse localization length in the Anderson model that
turns out to be valid both inside the energy band and at the band edge
(although it fails to reproduce the anomaly of the Lyapunov exponent at the
band centre).
The extended validity range of the expression~(\ref{lyap}) is a
remarkable feature, because the behaviour of the localization
length at the band edge is known to be anomalous~\cite{Der84,Izr98} and
has been previously derived with methods well distinct (and more
complicated) than the ones used to study the localization length inside
the band.
In the next two sections we will show how expression~(\ref{lyap})
reproduces the known formulae for the localization length inside the
energy band as well as in a neighbourhood of the band edge.

Before proceeding along this line, however, it is interesting to
notice that expressions very similar to those of Eqs.~(\ref{misinv})
and~(\ref{lyap}) have been obtained for a different but related model:
that of a particle in a one-dimensional random potential (see,
e.g.,~\cite{Lif88} and references therein). The problem is defined by
the continuous Schr\"{o}dinger equation
\begin{equation}
\psi''(x) + \left[ E - U(x) \right] \psi (x) = 0 \; ,
\label{schr}
\end{equation}
where $U(x)$ represents white-noise, i.e., a delta-correlated random
potential with zero mean:
\begin{eqnarray*}
\langle U(x) \rangle = 0 &; \;\;\;  & \langle U(x) U(x') \rangle
= 2 D \delta(x-x').
\end{eqnarray*}
This correspondence is not surprising on two grounds: first,
Eq.~(\ref{andmod}) is the discrete counterpart of the continuous
Schr\"{o}dinger equation~(\ref{schr}), and second, the stationary
equation~(\ref{schr}) is the formal analogue of the dynamical equation
for the kicked oscillator
\begin{displaymath}
\ddot{x}(t) + \left( \omega^{2} + \omega \xi(t) \right) x(t) = 0 \, ,
\end{displaymath}
which represents an equivalent form of the system~(\ref{dyneq}).

\section{Weak disorder expansion}
\label{wde}
The weak disorder case is defined by the condition~(\ref{weakdis}).
This condition implies that, except that at the band edge (i.e. for
$\omega T \rightarrow 0$), the parameter~(\ref{kappa}) must satisfy
the requirement $|k| \ll 1$.
In this section we analyse therefore the expansion of the Lyapunov
exponent~(\ref{lyap}) in the limit $k \rightarrow 0$. This corresponds
to studying the behaviour of the localization length inside the energy
band for the Anderson model~(\ref{andmod}) with a weak disorder.

Making use of expressions~(\ref{lyap}) and~(\ref{norm}), it is easy to
verify that for $k \rightarrow 0$ the Lyapunov exponent can be written
in the form
\begin{equation}
\lambda = \frac{\langle A_{n}^{2} \rangle}{4 T} \;
\frac{\displaystyle \sum_{n=0}^{\infty} (-1)^{n}
\frac{\Gamma \left( 3n + 3/2 \right)}{n!}
\left( \frac{k^{2}}{48} \right)^{n}}
{\displaystyle \sum_{n=0}^{\infty} (-1)^{n}
\frac{\Gamma \left( 3n + 1/2 \right)}{n!}
\left( \frac{k^{2}}{48} \right)^{n}} .
\label{expan}
\end{equation}
To the lowest order in $k$ this expression reduces to
\begin{equation}
\lambda = \frac{\langle A_{n}^{2} \rangle}{8 T} ,
\label{wklyap}
\end{equation}
which represents the basic approximation for the inverse localization
length in the weak disorder case.

Taking into account the relations~(\ref{param}) between the parameters
of the Anderson model~(\ref{andmod}) and those of the stochastic
oscillators~(\ref{kickosc}) and~(\ref{itoeq}), the variance $\langle
A_{n}^{2} \rangle$ that appears in formula~(\ref{wklyap}) can be
expressed as
\begin{displaymath}
\langle A_{n}^{2} \rangle = \frac{\langle \epsilon_{n}^{2} \rangle}
{1 - E^{2}/4} .
\end{displaymath}
When the distribution for the random site-energies $\epsilon_{n}$ is
the box distribution~(\ref{boxdis}), one can further write $\langle
\epsilon_{n}^{2} \rangle = W^{2}/12$; as a consequence,
expression~(\ref{wklyap}) takes the form
\begin{equation}
\lambda = \frac{1}{T} \frac{W^{2}}{96 \left( 1 - E^{2}/4 \right)} \;,
\label{stdlyap}
\end{equation}
which, for $T=1$, coincides with the well-known standard formula for
the inverse localization length in the Anderson model~\cite{Tho79}.

The extra-factor $1/T$ stems from definition~(\ref{lyapdef}) of the
Lyapunov exponent, which implies that $\lambda = \langle \dot{x}/x
\rangle$ has the dimension of a inverse time.
In order to have the correct physical dimension, therefore, $\lambda$
must be inversely proportional to a time parameter which, on the other
hand, must be a specific feature of the noise~(\ref{xi}), since that is
the physical origin of the orbit instability.
This requirement singles out the period $T$ between two kicks as the
only parameter which can endow $\lambda$ with the proper physical
dimension; the proportionality $\lambda \propto 1/T$ is thus fully
justified.

The expression~(\ref{stdlyap}) corresponds to the result derived by
Thouless using standard perturbation methods~\cite{Tho79}. As such,
formula~(\ref{stdlyap}) fails to reproduce the correct behaviour of
the Lyapunov exponent at the band centre, where the second order
perturbation theory of Thouless breaks down and an anomaly appears
which was first explained in Ref.~\cite{Kap81}.
This deviation of the inverse localization length from the behaviour
predicted by formula~(\ref{stdlyap}) is a resonance phenomenon, which can
be conveniently understood by considering the dynamics of the kicked
oscillator~(\ref{kickosc})~\cite{Izr98}.
In fact, the band centre corresponds to the case $\omega T = \pi/2$
and this equality can be interpreted as the condition that the
frequency $1/T$ of the kicks be exactly four times the frequency
$\omega/2\pi$ of the unperturbed oscillator. This generates a resonance
effect that manifests itself in a small but clear increase of the
localization length with respect to the value predicted by
formula~(\ref{stdlyap}) for $E=0$.
Once the origin of the anomaly at the band centre is explained in these
terms, it is not surprising that the model~(\ref{itoeq}) fails to
reproduce this feature, because it is obvious that the Wiener noise $W(t)$
cannot conveniently mimic the regularly time-spaced character of the
shot noise~(\ref{xi}).
One might then worry that the model~(\ref{itoeq}) provides an inadequate
description of the Anderson model whenever the period of the unperturbed
oscillator and that of the kicks stand in any rational ratio. This is not
the case, however, because the resonance effect at the band centre is
the only one that affects the localization length at the second order of
the perturbation theory~(\cite{Izr98}) and is thus of interest for the
present work.
For all the other `rational' values of the energy $E = 2 \cos \pi \alpha$
with $\alpha$ rational, the effect of the resonance on the Lyapunov
exponent can be seen only by going beyond the second order approximation
in the weak disorder expansion (see details in Ref.~\cite{Izr98}).

\section{The neighbourhood of the band edge}
Besides the $k \rightarrow 0$ limit considered in the previous section,
one can also study the behaviour of the Lyapunov exponent~(\ref{lyap})
in the complementary case $k \rightarrow \infty$.
Physically speaking, the limit $k \rightarrow \infty$ can be interpreted
in different ways, depending on the reference model. If one bears in mind
the kicked oscillator~(\ref{kickosc}), then taking the limit $k
\rightarrow \infty$ is tantamount to studying the case of a very strong
noise. More precisely, the condition $k \gg 1$ implies that the kicks
play a predominant role in the oscillator dynamics with respect to the
elastic force. Notice that this is not in contrast with the requirement
that each single kick be weak, as established by the
condition~(\ref{weakdis}). In fact, regardless of how weak the individual
kicks are, their collective effect can be arbitrarily enhanced by making
the interval $T$ between two successive kicks sufficiently shorter then
the fraction $\langle A_{n}^{2} \rangle / \omega$ of the period of the
unperturbed oscillator.

From the point of view of the Anderson model~(\ref{andmod}), the
analysis of the case $k \gg 1$ is equivalent to the study of the
localization length in a neighbourhood of the band edge. To understand
this point, one should consider that in the weak disorder case, as defined
by the relation~(\ref{weakdis}), the only way to fulfil the condition
$k = \langle A_{n}^{2} \rangle /(\omega T) \rightarrow \infty$ is to
have $\omega T \rightarrow 0$. Correspondingly, the energy $E = 2 \cos
(\omega T)$ must approach the limit $E \rightarrow 2^{-}$, i.e. the edge
of the band. Using relations~(\ref{param}), one can also express the
condition $k \gg 1$ in the significant form
\begin{displaymath}
2 - E \ll \langle \epsilon_{n}^{2} \rangle^{2/3} \;,
\end{displaymath}
which shows that the investigation of the case $k \gg 1$ corresponds
to studying the behaviour of the inverse localization length for energy
values which are close to the band edge on a distance scale set by the
fluctuations of the random site-potential.

With the physical meaning of the limit $k \rightarrow \infty$ clear in
mind, we can proceed to verify that Eq.~(\ref{lyap}) reproduces the
correct behaviour of the Lyapunov exponent in a neighbourhood of the
band edge. For this, it suffices to notice that the substitution
\begin{eqnarray}
k = \frac{\langle \epsilon_{n}^{2} \rangle}{\omega T \sin^{2} \left(
\omega T \right)} & \longrightarrow & k' =
\frac{\langle \epsilon_{n}^{2} \rangle}{\left( \omega T \right)^{3}}
\label{duek}
\end{eqnarray}
transforms formula~(\ref{lyap}) in the expression originally obtained
by Derrida and Gardner for the Lyapunov exponent at the band
edge~\cite{Der84}.
This implies that Derrida and Gardner's expression coincides with our
own for $\omega T \rightarrow 0$, since in this limit the difference
between parameters $k$ and $k'$ vanishes. The limit $\omega T
\rightarrow 0$, on the other hand, identifies the band edge case: this
proves that formula~(\ref{lyap}) is correct not only inside the energy
band (except that for $E=0$), but also for $E \rightarrow 2$.
The extended validity range of expression~(\ref{lyap}) is a relevant and
novel feature; indeed, to the best of our knowledge, no other formula
encompassing the {\em whole} energy band has been previously found for
the Lyapunov exponent in the Anderson model.

To conclude our discussion of the $k \rightarrow \infty$ limit, we observe
that in this case it may be appropriate to expand the integrals that
appear in expressions~(\ref{lyap}) and~(\ref{norm}) in series of the
inverse powers of $k$. One thus obtains
\begin{displaymath}
\lambda = \left( \frac{3}{4 k^{2}} \right)^{1/3}
\frac{\langle A_{n}^{2} \rangle}{T}
\frac{\displaystyle \sum_{n=0}^{\infty}
\frac{(-2\sqrt[3]{6})^{n}}{n!}
\Gamma \left( \frac{2n+3}{6} \right)
\left( k^{-1} \right)^{2n/3}}
{\displaystyle \sum_{n=0}^{\infty}
\frac{(-2\sqrt[3]{6})^{n}}{n!}
\Gamma \left( \frac{2n+1}{6} \right)
\left( k^{-1} \right)^{2n/3}} \;,
\end{displaymath}
which is the counterpart of the expansion~(\ref{expan}) of the
preceding section.
To the lowest order in $k^{-1}$ this expression reduces to
\begin{displaymath}
\lambda = \frac{1}{T} \frac{\sqrt[3]{6}}{2} \frac{\sqrt{\pi}}{ \Gamma
\left( 1/6 \right)} \left( \frac{\omega T}{\sin \left( \omega T \right)}
\right)^{2/3} \langle \epsilon_{n}^{2} \rangle^{1/3} \; ,
\end{displaymath}
so that for $E = 2$ (i.e., for $\omega T = 0$), the Lyapunov exponent
turns out to be
\begin{displaymath}
\lambda = \frac{1}{T} \frac{\sqrt[3]{6}}{2} \frac{\sqrt{\pi}}{ \Gamma
\left( 1/6 \right)} \langle \epsilon_{n}^{2} \rangle^{1/3} \; ,
\end{displaymath}
in perfect agreement with the result originally found in~\cite{Der84}
(see also~\cite{Izr98}).

\section{Dynamics of action-angle variables}

To gain further insight in the dynamics of the stochastic
oscillator~(\ref{itoeq}), it is useful to consider its description
in terms of action-angle variables.
To this end, we introduce the polar coordinates $(r,\theta)$ through
the standard relations
\begin{eqnarray}
r & = & \sqrt{x^{2} + p^{2}} \nonumber \\
\theta & = & \arctan \left( x/p \right)
\label{polar}
\end{eqnarray}
Starting from this definition and Eqs.~(\ref{itoeq}), it is
straightforward to obtain the following pair of stochastic It\^{o}
equations
\begin{eqnarray}
dr & = & \frac{\langle A_{n}^{2} \rangle}{2T}
r \sin^{4} \theta \; dt - \sqrt{\frac{\langle A_{n}^{2} \rangle}{T}}
r \sin \theta \, \cos \theta \; dW(t)
\label{action} \\
d \theta & = & \left( \omega + \frac{\langle A_{n}^{2}
\rangle}{T} \cos \theta \sin^{3} \theta \right) dt +
\sqrt{\frac{\langle A_{n}^{2} \rangle}{T}} \sin^{2} \theta  \; dW(t)
\label{angle}
\end{eqnarray}
A simple examination of Eqs.~(\ref{action}) and~(\ref{angle}) reveals
that, while the action variable is coupled to the angular one, the
latter evolves in a independent way. Therefore, one can determine the
probability distribution for the variable $\theta$ by solving the
Fokker-Planck equation associated to the It\^{o}
equation~(\ref{angle})~\cite{Gar83,Kam92}, i.e.,
\begin{eqnarray}
\frac{\partial \rho}{\partial t} \left( \theta,t \right) & = &
-\frac{\partial}{\partial \theta} \left[ \left( \omega + \frac{\langle
A_{n}^{2} \rangle}{T} \cos \theta \, \sin^{3} \theta \right) \rho \left(
\theta, t \right) \right] \nonumber \\
& & + \frac{\langle A_{n}^{2} \rangle}{2T}
\frac{\partial^{2}}{\partial \theta^{2}} \left[ \left( \sin^{4} \theta
\right) \, \rho \left( \theta, t \right) \right] \; .
\label{pofkpl}
\end{eqnarray}

In geometrical terms, the introduction of the angular variable $\theta$ is
equivalent to the projection of the point $(x(t),p(t))$, representative of
the stochastic oscillator~(\ref{itoeq}) in the phase space, onto the
unit circle by the relation~(\ref{polar}). A glance at the Fokker-Planck
equation~(\ref{pofkpl}) shows that the ensuing motion of the projected
point $\theta(t)$ is the combination of a drift with a noise-induced
diffusion. In the Fokker-Planck equation~(\ref{pofkpl}) the diffusion term
vanishes for $\theta = 0$ and $\theta= \pi$, but the drift coefficient is
not zero at these points (unless $\omega \rightarrow 0$), so that it is
reasonable to expect the invariant measure $\rho (\theta)$ to be finite
and non-vanishing at any point of the interval $[0:2\pi]$.
Things are different when $\omega \rightarrow 0$, i.e., at the band
edge, for in this case both the drift and the diffusion coefficient
tend to vanish in a neighbourhood of $\theta =0$ and $\theta = \pi$.
Therefore, for $\omega \rightarrow 0$, the projection of the
representative point onto the unit circle tends to stick at the critical
points $\theta =0$ and $\theta = \pi$; as a consequence, one can expect
the invariant measure $\rho (\theta)$ to be sharply peaked around these
two values of $\theta$ and almost zero everywhere else.
This peculiar behaviour for $\omega \rightarrow 0$ gives an intuitive
feeling for the origin of the anomalous scaling of the Lyapunov exponent
with the noise strength at the band edge.

The stationary solutions of~(\ref{pofkpl}) are periodic in $\theta$ with
period $\pi$. In the interval $[0:\pi]$ the solution $\rho(\theta)$ which
is bounded as $\theta \rightarrow \pi$ and is normalized with the condition
$\int_{0}^{2\pi} \rho(\theta) \, d\theta = 1$ can be expressed in the
integral form
\begin{equation}
\rho \left( \theta \right) = \frac{1}{2N \, \sin^{2}\theta}
\int_{\theta}^{\pi} \frac{dx}{\sin^{2} x} \, \exp \left\{ \Phi \left(
\cot x \right) -\Phi \left( \cot \theta \right) \right\} \; ,
\label{poldis}
\end{equation}
where $\Phi$ is the function~(\ref{phi}) and $N$ is the normalization
constant~(\ref{norm}); outside the interval $[0:\pi]$, the probability
$\rho (\theta)$ is defined through the periodicity condition
$\rho (\theta + k \pi) = \rho (\theta)$ with $k$ integer.

Notice that at the points $\theta = 0$ and $\theta = \pi$ the invariant
measure $\rho(\theta)$ takes the value $\rho (0) = \rho (\pi) = k/4N$.
Expanding expression~(\ref{norm}) in power series of $k$ and $k^{-1}$, it
is possible to show that in the opposite limits $k \rightarrow 0$ and
$k \rightarrow \infty$ one has
\begin{eqnarray*}
\rho(0) & = & \rho(\pi) = \frac{k}{4N(k)} \\
& = & \left\{
\begin{array}{ll}
\frac{1}{2\pi} + O\left( k^{2} \right) & \mbox{for $k \rightarrow 0$}\\
\frac{3}{6^{1/6} (8\pi)^{1/2} \Gamma(1/6)} \, k^{1/3} + O\left(1/k^{2/3}
\right) & \mbox{for $k \rightarrow \infty$}
\end{array}
\right. \; .
\end{eqnarray*}
These expansions show that, upon increasing $k$, the probability
distribution~(\ref{poldis}) does become more and more sharply peaked
at the points $\theta = 0$ and $\theta = \pi$; the analysis of the
function $\rho(\theta)$ thus confirms the singular behaviour of the
invariant measure that was suggested from physical considerations on
the dynamics of the stochastic variable $\theta(t)$.

Using~(\ref{poldis}) one can easily compute the Lyapunov exponent
as defined by the expression~(\ref{lyapdef}). To this end, it suffices
to notice that in polar coordinates the Ricatti variable takes the form
$z = \omega \cot \theta$: one can therefore evaluate the Lyapunov
exponent as
\begin{eqnarray*}
\lambda & = & \omega \langle \cot \theta \rangle = 2\omega \int_{0}^{\pi}
d\theta \, \rho \left( \theta \right) \cot \theta =\\
& = & \frac{\omega}{N}
\int_{0}^{\pi} \frac{d \theta}{\sin^{2}\theta} \cot \theta \,
\exp \left\{ -\Phi \left( \cot \theta \right) \right\} \\
& \cdot &
\int_{\theta}^{\pi} \frac{dx}{\sin^{2} x} \, \exp \left\{ \Phi \left(
\cot x \right) \right\} \, .
\end{eqnarray*}
With a change of variables this expression can be reduced to the
form~(\ref{averag}), so that one eventually recovers the
result~(\ref{lyap}) for the Lyapunov exponent.
This is not surprising, since the distributions~(\ref{misinv})
and~(\ref{poldis}) are strictly related; indeed one has
\begin{displaymath}
\rho(\theta) = \frac{1}{2} \, p\left( z \left( \theta \right) \right)
\left| \frac{dz}{d\theta} \right| \, ,
\end{displaymath}
where the factor $1/2$ derives from the normalization conditions chosen
for the two distributions. From a geometrical point of view, the passage
from the process $z(t)$ to $\theta(t)$ is equivalent to projecting the real
line $-\infty < z(t) < +\infty$ onto the unit circle $0 < \theta(t) < \pi$
via the relation $\theta(t) = \cot^{-1} \left( z/\omega \right)$.

\section{Acknowledgments}

FMI gratefully acknowledges support by the CONACyT (Mexico) Grants
No. 26163-E and No. 28626-E.

\end{multicols}
\end{document}